\documentclass[%
 reprint,
superscriptaddress,
showpacs,preprintnumbers,
 amsmath,amssymb,
 aps,
floatfix,
]{revtex4-1}
\usepackage{mathtools}
\usepackage{graphicx}
\usepackage{dcolumn}
\usepackage{bm}
\usepackage{hhline}
\usepackage{color,soul}
\usepackage{float}

\definecolor{green}{rgb}{0.0, 0.5, 0.0}

\begin{document}


\title{Wrinkles, folds and ripplocations: unusual deformation structures of 
confined elastic sheets at non-zero temperatures}
\author{Debankur Das}
\affiliation{%
Tata Institute for Fundamental Research,  Centre for Interdisciplinary Sciences, 36/P Gopanapally, Hyderabad 500107,  India.
}%
\author{J\"urgen Horbach}
\affiliation{Institut f\"ur Theoretische Physik II: Weiche Materie, Heinrich 
Heine-Universit\"at D\"usseldorf, Universit\"atsstra{\ss}e 1, 40225 D\"usseldorf, 
Germany}
\author{Peter Sollich}
\affiliation{%
King's College London, Department of Mathematics, Strand, London WC2R 2LS, UK. Current address: University of G\"ottingen, Institute for Theoretical Physics, 37077 G\"ottingen, Germany 
}%
\author{Tanusri Saha-Dasgupta}
\affiliation{Department of Condensed Matter Physics and Materials Science,
S.N. Bose National Centre for Basic Sciences, Kolkata 700098, India.}
\author{Surajit Sengupta}
\affiliation{%
 Tata Institute for Fundamental Research,  Centre for Interdisciplinary Sciences, 36/P Gopanapally, Hyderabad 500107,  India.
}%

\date{\today}
\begin{abstract}
We study the deformation of a fluctuating crystalline sheet confined
between two flat rigid walls as a simple model for layered solids
where bonds among atoms {\it within} the same layer are much stronger
than those {\it between} layers.  When subjected to sufficiently
high loads in an appropriate geometry, these solids deform and fail
in unconventional ways. Recent experiments suggest that configurations
named {\it ripplocations}, where a layer folds backwards over itself,
are involved. These structures are distinct and separated by large
free energy barriers from smooth {\it ripples} of the atomic layers
that are always present at any non-zero temperature. We use Monte
Carlo simulation in combination with an umbrella 
sampling technique to obtain conditions under which such structures form 
and study
their specific experimental signatures.
\end{abstract}
\maketitle
\section{\label{sec:level1} Introduction}
The current understanding of the mechanical
response of materials to large external stress~\cite{CL, Landau,
rob} is mostly based on ideas applicable to simple close-packed
solids \cite{CL}.  Prevalent theories of irreversible deformation,
therefore, invariably involve the nucleation and motion of lattice
defects such as dislocations~\cite{hirth,nabarro}.  Recently, it
has been observed that spatially confined flexible membranes deform
by reorganizing their morphology to form hierarchical structures
of great complexity~\cite{hierarchy}. Wrinkles, ripples,
folds~\cite{Milner,Cerda-Maha,Zhang,Brau,Li-wrinkle}, as well as
higher order structures like pleats~\cite{sas4,sas7} are ubiquitous
both in nature and in emergent technology. They
have been observed in biological tissues~\cite{spectrin1,spectrin2},
polymer sheets~\cite{muthu}, and many other low-dimensional materials
involving flexible sheets and membranes~\cite{origami, kirigami1,
kirigami2, irvine, grason1, grason2, grason3, narayanan1, narayanan2,
flat-fold1, flat-fold2}. Similar structures,
named ripplocations, have been introduced to describe new kink-like
deformation mechanisms ~\cite{frank}, associated with the buckling
of surface layers in response to mechanical loading of van der
Waals-layered solids such as MoS$_2$ or the MAX family of solids
like Ti$_3$SiC$_2$, graphene etc.~\cite{barsoum2003, 1barsoum2003,
barsoum2004, 1barsoum2004, barsoum2005, barsoum2013, kushima2015,
gruber2016, Tucker, Freiburg}. Ripplocations are structurally
distinct from conventional dislocations in bulk crystals \cite{rob,
nabarro, hirth}. Such patterns are prevalent in many types of deformed layered materials spanning more than 13 orders of magnitude
in scale~\cite{Barsoum-Zhao}, including massive
geological formations such as phyllosilicates in the lithosphere~\cite{Aslin2019}.

\begin{figure}[ht]
\includegraphics[scale=0.24]{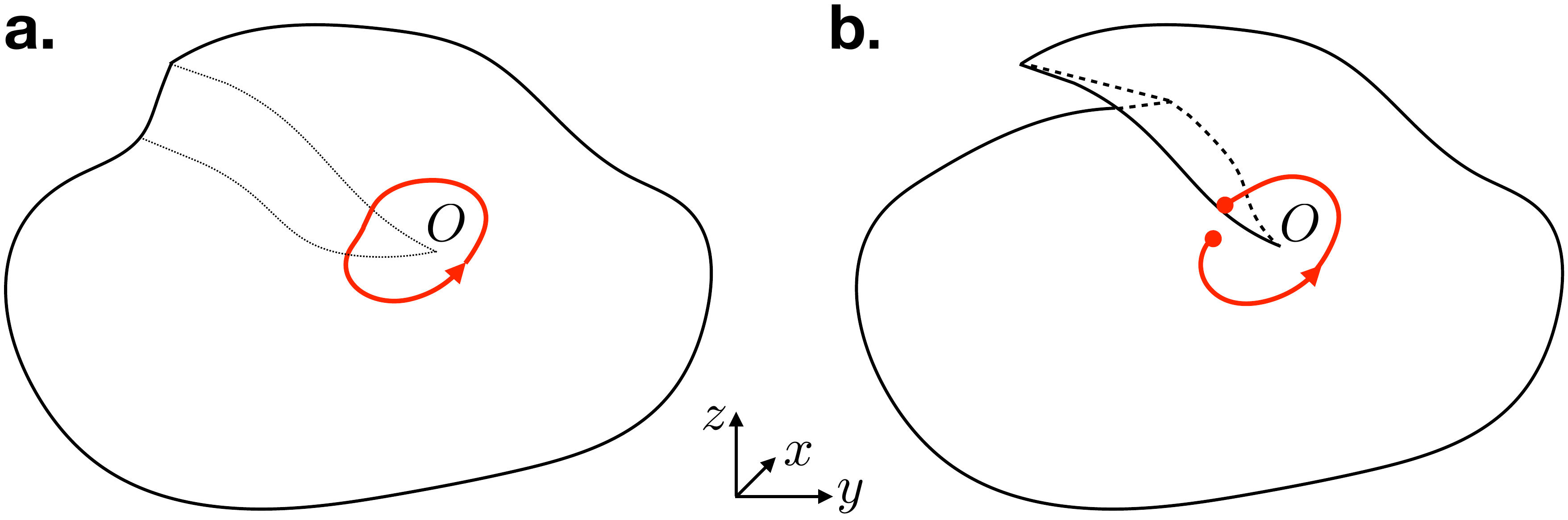}
\caption{Schematic figure showing a ripple {\bf a.} and a ripplocation
{\bf b.} in a layered solid. The red circles around the point $O$ with arrows show Burgers circuits on the layer. In
a ripple, such a circuit returns to the same point on the layer
whereas if a ripplocation is present, the circuit returns to a point
displaced in the direction perpendicular to the layer. Assuming that the boundary of the sheet is single
valued, every point, $O$, for a ripplocation is associated with a
corresponding  point $O'$ (not shown for clarity) where an identical
circuit produces an equal displacement of opposite sign.}
\label{schem1}
\end{figure}

Small compressive strains in layered materials
result in the formation of smooth undulations, known as ripples or {\it
wrinkles}. These undulations are associated with a broad distribution of
strain energy.  At larger compression, the strain energy can be localized,
leading to structures with sharp {\it folds}. Thus, there can be a {\it
wrinkle-to-fold} transition with increasing compressive strain.  At zero
temperature, $T=0$, this behavior can be described by the F\"oppl-von
K\'arm\'an equations~\cite{Landau} that in general cannot be solved
analytically. Nevertheless, approximate theories of the wrinkle-to-fold
instability have been derived~\cite{Cerda-Maha}. Further compression leads
to higher-order deformation patterns where atomic layers glide relative
to each other without breaking the in-plane bonds and producing a pleat
(see Fig.\ref{schem1}) or {\it ripplocation}. These structures involve
large and singular deformations of flat sheets and are intractable within
existing elasticity theory.

Although the $T=0$ energetics of system-spanning ripplocation has
been studied~\cite{Freiburg,kushima2015}, many questions remain
unanswered. A completely open issue is the mechanism
for the formation of ripplocations at finite temperatures, $T \neq
0$.  Also the following questions about the emergence of ripplocations
are not well understood:  What are the microscopic precursors and
intermediate states that give rise to these higher order structures?
How are such incipient structures to be distinguished from thermally
driven random height fluctuations? What is the typical free energy
barrier involved in their formation? What are the typical mechanical
signatures of ripplocation formation?

In this paper, we address these issues in the context of a thin
fluctuating sheet modeled as a network of connected vertices ordered
in a triangular lattice and confined between two rigid layers. Our model system does not correspond to any particular material. Rather, we provide answers to some of the many
questions raised above depending on physical parameters such as the
temperature, the stiffness of the layers, the intra-layer mechanical
coupling etc., which vary from material to material.  Our calculations
explicitly take into account the effect of finite temperature $T >
0$ and therefore represent a definite advance on earlier work.  We
show that ripplocation like structures, which are the generalization
of the purely two-dimensional (2D) pleats studied earlier~\cite{sas4,
sas7}, readily form following a phase transition from an essentially
flat sheet containing at most thermally generated ripples.  These
phases co-exist at a first-order boundary. The free energy barrier
between these phases at co-existence is large but reduces as the
solid is deformed. For certain choices of parameters the ripplocated
phase remains metastable at all values of strain.

The difference between normal, smooth fluctuations of the height
of a sheet, which has been variously called a wrinkle, ripple or
fold, and one that comprises a pleat or ripplocation is explained
schematically in Fig.\ref{schem1}. A smooth wrinkle (c.f.
Fig.\ref{schem1}{\bf a}), is always representable as a single
valued function $z(x,y)$. Traveling along any closed loop on the
surface always ends in a return to the starting
point anywhere on this surface. The scale of these smooth ripples
may vary. We denote small random fluctuations of height as wrinkles
and large collective height fluctuations are called folds.

In contrast, a ripplocation or pleat (see Fig.\ref{schem1}{\bf b})
involves a multivalued height function $z(x,y)$
and there is the existence of points $O$ for which closed loops enclosing them are finally always displaced by an
amount $\Delta z$ from the starting point perpendicular to the
plane. Now imagine a pair of such singular points
$O$ and $O'$ adjacent to one another with displacements $\pm \Delta
z$, of opposite signs. These can be viewed as the analogs of a
dislocation dipole in a 2D solid~\cite{CL,rob}. However, in dislocation dipoles, the displacement, known as the
Burgers vector, is within the plane and along the line joining $O$
and $O'$. We show later that system spanning ripplocations form
when $O$ and $O'$ separate in the presence of external loads.

In order to study the statistical mechanics of ripplocations, we
need to construct an appropriate collective variable to be used as
a reaction co-ordinate in order to obtain the
free energy surface as well as free energy barriers. To this end,
we employ a quantity for the measure of non-affine displacements
that was first introduced in the study of mechanical deformations
in glasses~\cite{falk}. Subsequently, this quantity has been
generalized~\cite{sas1} and used to investigate defects in
crystals~\cite{popli}, pleats in permanently bonded
networks~\cite{sas4,sas7}, the origin of rigidity of crystalline
solids~\cite{pnas}, and biologically important conformation changes
in proteins~\cite{protein}.

The rest of the paper is organized as follows. In the next section
(Section II), we introduce our model for the fluctuating, confined
two-dimensional sheet, define the collective
variable to measure non-affine displacements and give the details
of our simulation method. In Section III, we discuss our results
starting with a re-examination of the ripple-to-fold instability
in our model, followed by a description of the equilibrium transition
from ripple to ripplocation at non-zero temperatures and the
presentation of the computed equilibrium phase diagram. Then,
intermediate structures that arise during nucleation of the ripplocated
phase are analyzed. Finally, we conclude the paper in Section IV
by discussing the possible experimental implications and future
directions of research.

\section{\label{sec:level1}Model and Methods}
\subsection{\label{sec:level2} The fluctuating sheet as a model network}
\begin{figure}[t]
\includegraphics[scale=0.37]{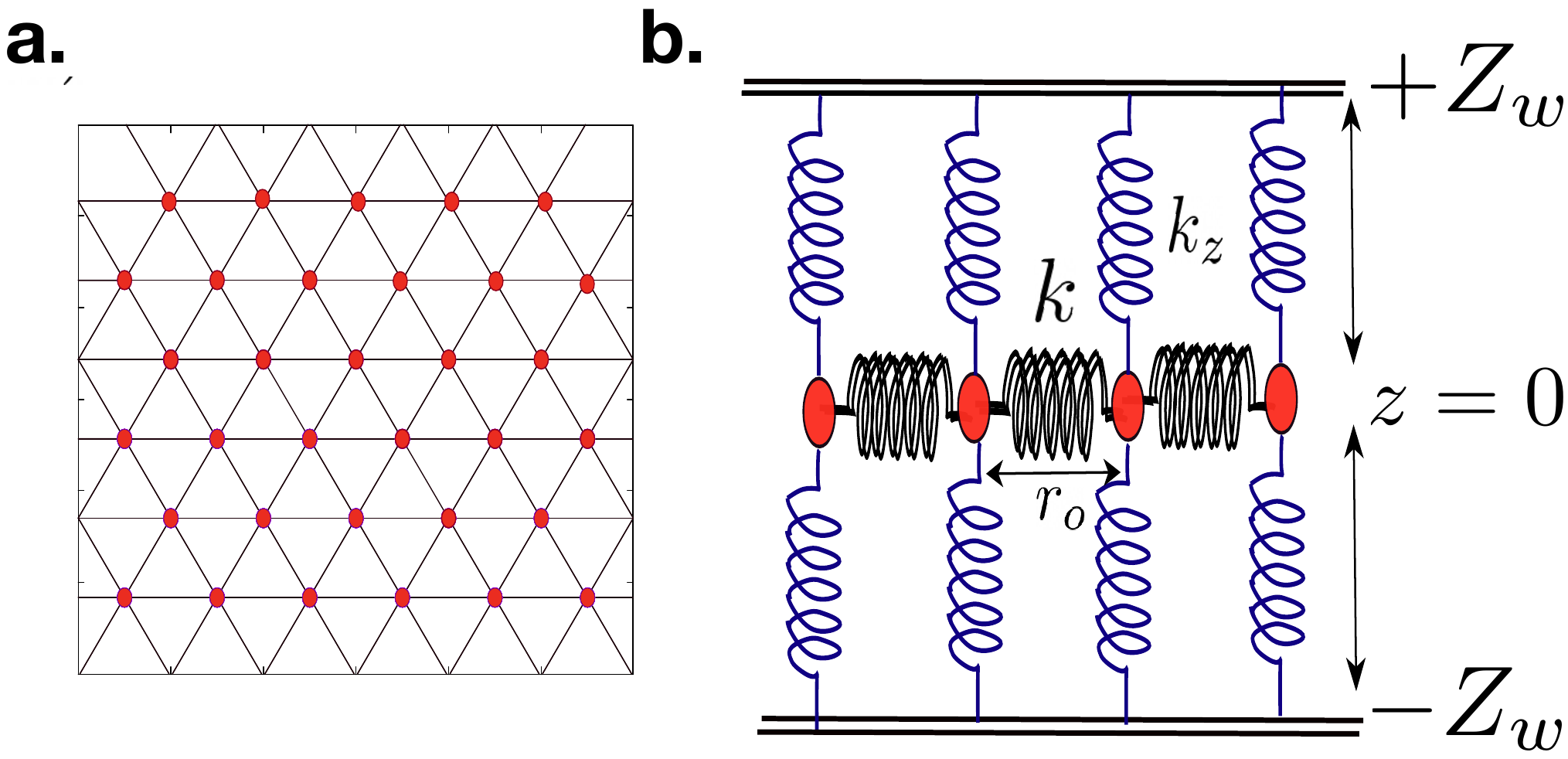}
\caption{Schematic of the model network confined by two rigid walls.
{\bf a.} {\it Top view} showing the initial triangular lattice in
the $xy$ plane. The red colored beads indicate the initial positions
of the particles. {\bf b.} {\it Lateral view} showing the springs
({\it blue}) connecting the particles with the confining walls.
\label{model}}
\end{figure}
We model the fluctuating sheet as a network of vertices connected
by elastic bonds. The model consists of $N$ particles, interacting via a harmonic potential with respect to a 2D
reference network structure (see below) and a repulsive Weeks-Chandler-Andersen 
(WCA)~\cite{CL} potential $u_{\rm WCA}(r)$. The latter potential is defined by 
\begin{eqnarray}
\label{eq_wca}
u_{\rm WCA}(r) &=& 
\begin{cases}
4e\left[\left(\frac{\sigma_0}{r}\right)^{12}-
   \left(\frac{\sigma_0}{r}\right)^6\right] + e & \text{if }  |r| \leq r_c\\
0              & \text{otherwise},
\end{cases}
\end{eqnarray}
with $r$ the distance between a pair of particles. The radius $r_c$
is set to $r_c=2^{1/6}\sigma_0$.  So the potential, defined by
Eq.~(\ref{eq_wca}), is a Lennard-Jones potential that is cut off
and shifted to zero at its minimum.

The initial positions of the network are at ${{\bf
R}_{i}}$ ($i=1,\ldots,N$), corresponding to an ideal triangular
reference lattice in the $xy$ plane (Fig.\ref{model}{\bf a}).  The
particles are connected by harmonic bonds of stiffness $k$ and
length $R_0=|{\bf R}_i-{\bf R}_j|$ (with ${\bf R}_i$ and ${\bf R}_j$
being reference lattice vectors of two adjacent particles $i$ and
$j$, respectively).  In $z$ direction, the network is confined by
two parallel walls that are at a distance $\pm Z_w$ on either side of
the triangular lattice plane. So the Hamiltonian of the harmonic
network model is given by
\begin{equation}
\mathcal{H} = \mathcal{H}_0 + \mathcal{H}_{\rm WCA} 
              + \mathcal{H}^{\rm wall}
\label{eq_ham}
\end{equation}
Here, $\mathcal{H}_0$ consists of the kinetic
energy and the potential energy due to the harmonic bonds,
\begin{equation}
\mathcal{H}_{0} = \sum_{i=1}^N \frac{{\bf p}_{i}^2}{2m} + 
\frac{k}{2} \sum_{i=1}^N \sum_{j\in \Omega, j < i} 
(|{\bf r}_j -{\bf r}_i| - |{\bf R}_j - {\bf R}_i|)^2 
\end{equation}
with $m$ the mass of a particle, ${\bf p}_i$ the
momentum of particle $i$, and ${\bf r}_i$ its instantaneous position.
$\Omega$ signifies the interaction volume, which for each node $i$
comprises all its nearest neighbours $j$ on the triangular lattice.
The term $\mathcal{H}_{\rm WCA}$ in Eq.~(\ref{eq_ham}) is the potential 
energy due to the repulsive WCA interactions,
\begin{equation}
\mathcal{H}_{\rm WCA} = \sum_{i=1}^N \sum_{j>i} u_{\rm WCA}(r_{ij}),
\end{equation}
where $r_{ij}=|{\bf r}_j-{\bf r}_j|$ and $\sum_{i=1}^N
\sum_{j>i}$ denotes the double sum over all $\frac{1}{2}N(N-1)$
particle pairs.

The term $\mathcal{H}^{\rm wall}$ describes the
interaction of the particles with the confining flat walls. The particles are connected in $z$ direction to both the walls by harmonic springs of unstretched length $Z_o$ and stiffness $k_z$. In addition, there are repulsive WCA interactions between the particles and the
walls in $z$ direction. Thus, the particle-wall interactions are
given by
\begin{eqnarray*}
\mathcal{H}^{\rm wall} &=& \mathcal{H}^{\rm wall}_{0} + 
\mathcal{H}^{\rm wall}_{\rm WCA}\\
\mathcal{H}^{\rm wall}_{0} &=& 
\sum_i \frac{k_z}{2}\left\{ (|Z_i-Z_w|-Z_0)^2 \right.\\
&& \left.+ (|Z_i + Z_w|-Z_0)^2 \right\}\\
\mathcal{H}^{\rm wall}_{\rm WCA} &=& 
\sum_i [u_{\rm WCA}^{\rm w}(Z_w-Z_i) + u_{\rm WCA}^{\rm w}(Z_i+Z_w)]\\
u_{\rm WCA}^{\rm w}(r) &=& \begin{cases}
4e_{\rm w} \left[\left( \frac{\sigma_{\rm w}}{r}\right)^{12}
                -\left( \frac{\sigma_{\rm w}}{r}\right)^6 \right] 
+ e_{\rm w} & \text{if} ~|r| \leq Z_c \\
    0              & \text{otherwise}
\end{cases}
\end{eqnarray*} 
with $Z_c = 2^{1/6}\sigma_{\rm w}$. That the particles cannot overlap with each other or with the walls is ensured by finite values of
$\sigma_0$ and $\sigma_{\rm w}$. We chose the unit of length as $R_0$ and  the unit of energy as $e$. For large values of $\sigma_0 (\simeq R_0)$, slip dislocations can occur on application of compression. These are not of interest here so we present results for $\sigma_0 = 0.1 R_0$ and $e = 1.0$. The qualitative results seem to be largely insensitive to the value of
$\sigma_0$ as long as $\sigma_0 \ll R_0$. For our  calculations, we choose $Z_0 = 2.4,R_0 =1.0$ so that the ratio $Z_0/R_0$ is
similar to the ratio of the typical in-plane interparticle distance
to the distance between layers in a solid such as graphite~\cite{CL}.
We have set the parameters $m=1$,$k=1$,$\sigma_{\rm w} = \sigma_0$ and $e_{\rm
w} = e$.

\subsection{\label{sec:level2} The collective order parameter for ripplocations}
The primary ingredient for our analysis is contained in a paper by
Ganguly {\it et al.}~\cite{sas1}, which shows that any displacement
of atoms away from a reference position can be classified into two
mutually orthogonal subspaces using a projection operator ${\mathsf
P}(\{{\bf R}\})$ which depends only on the set of lattice vectors
$\{{\bf R}\}$ defining a suitably chosen reference configuration.
These subspaces may be classified as ``affine'' and ``non-affine''
depending on whether or not the instantaneous
particle positions, $\{{\bf r}\}$, of the displaced structures can
be represented as an affine deformation (homogeneous strain) of the
reference configuration. The affine subspace may be parameterized
by the local deformation tensor ${\mathsf D}$ while the non-affine
part of the displacements is parameterized by a local scalar $\chi$.
The latter is the least square error made by fitting $\{{\bf r}\}$
to a ``best fit'' ${\mathsf D}$. Various thermodynamic quantities
such as the ensemble average $\langle \chi \rangle$ and the
spatio-temporal correlation functions $\langle \chi(0,0)\chi({\bf
R},t)\rangle$ can be obtained analytically~\cite{sas1,sas2}. The
spatial average $X = N^{-1} \sum_{i} \chi({\bf R}_i)$ behaves as a
thermodynamic variable with a conjugate field $h_X$~\cite{sas2}.

We study the network in the presence of both $h_X$ and an externally
imposed strain $\epsilon_d$. Specifically, we have a rectangular
box whose dimensions are commensurate with the triangular lattice.
The strain is implemented by expanding the simulation box along the
$y$-direction and compressing it along the $x$- and $z$-direction
by the same fractional amount $\epsilon_d$ while conserving volume
to linear order. Tuning $h_X$ can either suppress or enhance lattice
defects and this can be used to study the deformation of solids.
Indeed, we have shown~\cite{pnas} that an initially ideal (defect-free)
2D crystal when deformed becomes metastable for infinitesimal
deformation and tends to decay into the stable state where stress
is eliminated by slipping of crystalline planes (lines in 2D). This
is the dynamical consequence of an underlying first-order phase
transition as a function of both $h_X$ and deformation. Although
it is difficult (though not entirely impossible, see~\cite{poplitrap})
to realize $h_X$ experimentally, this expansion of parameter space
provides many physical insights, with experimentally realizable
consequences being recovered in the $h_X \to 0$ limit. We therefore
follow a strategy similar to that employed in Refs.~\cite{sas4,sas7,pnas}.

\subsection{\label{sec:level2} Successive umbrella sampling}
To study the ripple-to-ripplocation transition as a function of
$h_X$ and deformation at non-zero temperatures, an efficient
computational scheme that is able to access structures with
ripplocations starting from a flat sheet is required. While ripples
are always present at non-zero temperatures, we show later that
transition probabilities between ripples and ripplocations are
exponentially small due to large barriers. One method that produces
satisfactory results in this case is successive umbrella sampling
Monte Carlo (SUS-MC)~\cite{SUS}, which has been used previously to
study pleats in strictly two dimensional networks~\cite{sas4,sas7}.

To implement SUS-MC for our system, we divide the range of the
reaction coordinate $X$ into small windows and sample configurations
generated by Metropolis Monte Carlo ~\cite{binder, ums} in each
window, keeping the system restricted to the chosen window for a
predefined number of MC cycles. Beginning with $X = 0$, histograms
are recorded to keep track of accepted MC moves and how often the
system tries to leave a window via its left or right boundary. The
probability distribution $P(X)$ can then be computed from these
histograms. Further details of the procedure can be found in our
earlier work~\cite{sas4,sas7,pnas}. The computational effort needed
for implementation of SUS-MC is nonetheless substantial and grows
with $N$. This restricts the system size. By studying the finite
size effects at $T > 0$, conclusions can be drawn by extrapolating
to the thermodynamic limit $N \to \infty$. In this work, we present
results mostly for $ N = 900 $ vertices, mentioning consequences
of finite size later wherever appropriate. We used $ \beta = 1/(k_{\rm
B}T)=200 $ throughout; results at other temperatures are qualitatively
similar. We divide the range $0 < X < 0.4$ into $800$ windows to
obtain sufficient resolution. For sufficiently averaged $P(X)$
values, $2\times10^8$ trial moves are required in each window. Once
we obtain $P(X)$ for one particular combination of $h_X$ and
$\epsilon_d$ value, we use the histogram reweighting technique to
determine $P(X)$ at any other $h_X$~\cite{borgs90, ferrenberg88}.

\section{\label{sec:level1} Wrinkle, fold and ripplocation 
transitions\protect\\}
When a flexible sheet is compressed, at $T=0$, two specific kinds
of transitions are observed. Firstly, small undulations, or wrinkles
appear, which on further compression give rise to
folds~\cite{Milner,Cerda-Maha,Zhang,Brau,Li-wrinkle}. As explained
in the Introduction, the height variable in both wrinkles and folds
is single-valued and we collectively call these configurations
ripples. Under certain circumstances, such rippled phases can produce
multi-valued height fluctuations, which we will refer to as
ripplocations or pleats. We describe below first the wrinkle to
fold transition at $T > 0$, and then consider and analyze the 
transition to the ripplocated phase.

\subsection{\label{sec:level2} Finite temperature wrinkle to fold instability}
\begin{figure}[t]
\centering
\includegraphics[scale=0.3]{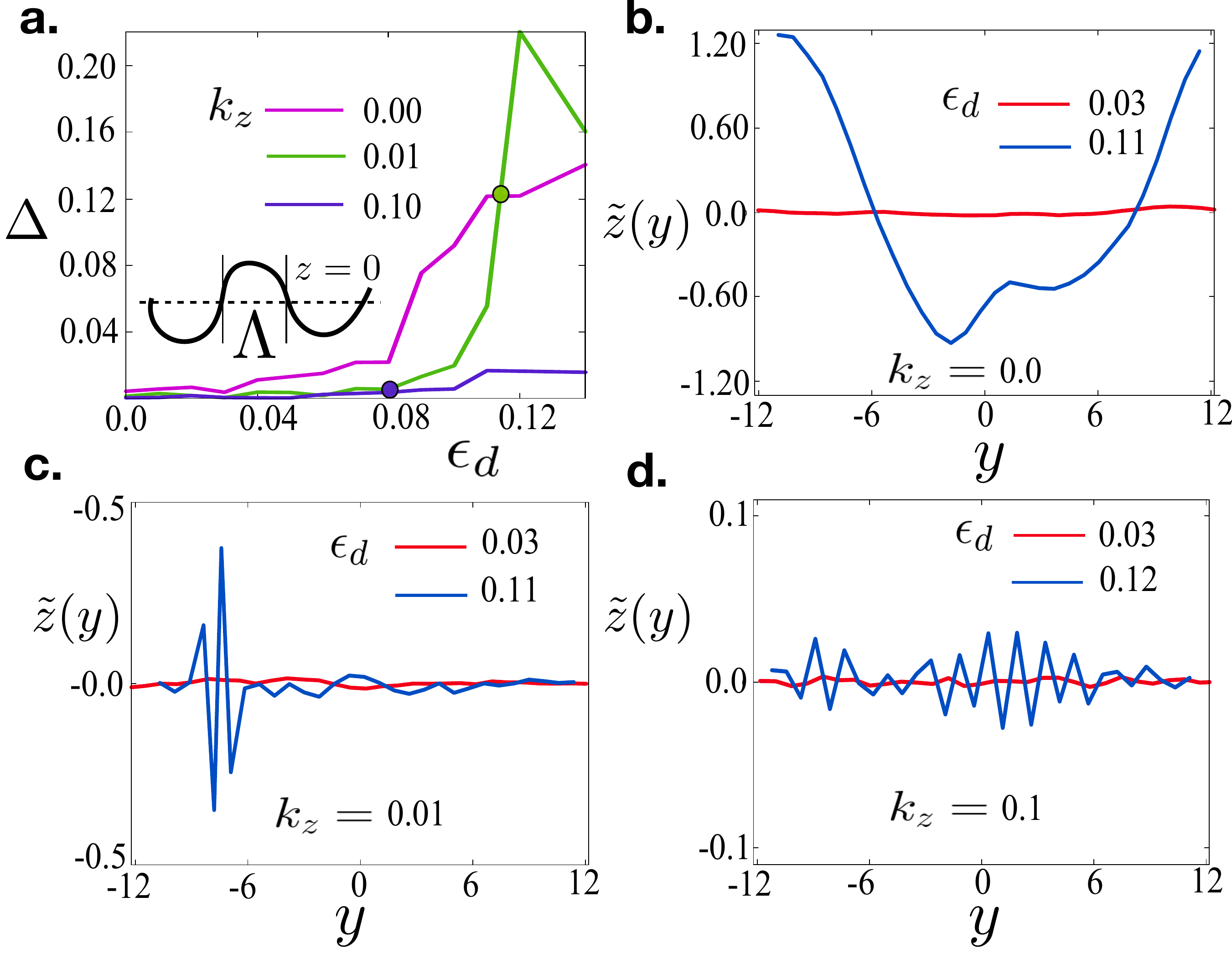}
\caption{{\bf a.} Plot of $\langle \Delta \rangle$ vs.~$\epsilon_d$
at $R_1$ (see text) obtained from SUS-MC of a $30 \times 30$ lattice
with $k_z = 0.0, 0.01, 0.1$. The inset shows the definition of the length
scale $\Lambda$. Note that $\langle \Delta \rangle$ undergoes a
finite jump at $\epsilon_d \approx 0.08$ for $k_z = 0.0$ and at
$\epsilon_d \approx 0.1$ at $k_z = 0.01$ signifying
a wrinkle-to-fold transition. Note that for $k_z = 0.1$, such a
transition is not observed. The green and blue circles indicate the
$\epsilon_d$ at the ripple-ripplocation transition for $k_z = 0.01$
and $k_z = 0.1$, respectively.The $x$-averaged $z$-value $\tilde{z}(y)$
at each $y$ is plotted before and after the wrinkle-fold transition
for {\bf b.} $k_z = 0.0$ and {\bf c.} $k_z = 0.01$ {\bf d.} Plot
of $\tilde{z}(y)$ vs.~$y$ for $k_z = 0.1$ and $\epsilon_d = 0.03, 0.12$.
\label{wrinkle}}
\end{figure}

The elastic instability that gives rise to the wrinkle-to-fold
transition persists at $T > 0$. However, due to
the thermal motion at $T>0$, it becomes difficult to
distinguish between thermal height fluctuations and elastic buckling.
Nevertheless, we define the quantity $\Delta = {\rm
max}[|\tilde{z}(y)|]/\Lambda$, where $\tilde{z}(y)$ is the $x$-averaged
value of the height $z$ at each position $y$ along the direction
of compression. We define $\Lambda$ as a typical width of height
fluctuations (see Fig.\ref{wrinkle}, inset). We calculate the
thermal and spatial average $\langle \Delta \rangle$ from configurations
obtained from our SUS-MC simulations. We ensure that these
configurations contain only the ripple phase, i.e.\ the height
function is always single-valued on average. In the next section
(Section~\ref{riptoriplo}) we show that the
probability distribution $P(X)$ always has a maximum (or the effective
dimensionless free energy $-\ln P(X)$ a minimum) at small $X$ (denoted by $R_1$), which corresponds to the
ripple phase.  Here, we analyze configurations that belong to this
small $X$-minimum as extracted from our SUS-MC calculations.

\begin{figure*}[ht]
\includegraphics[scale=0.35]{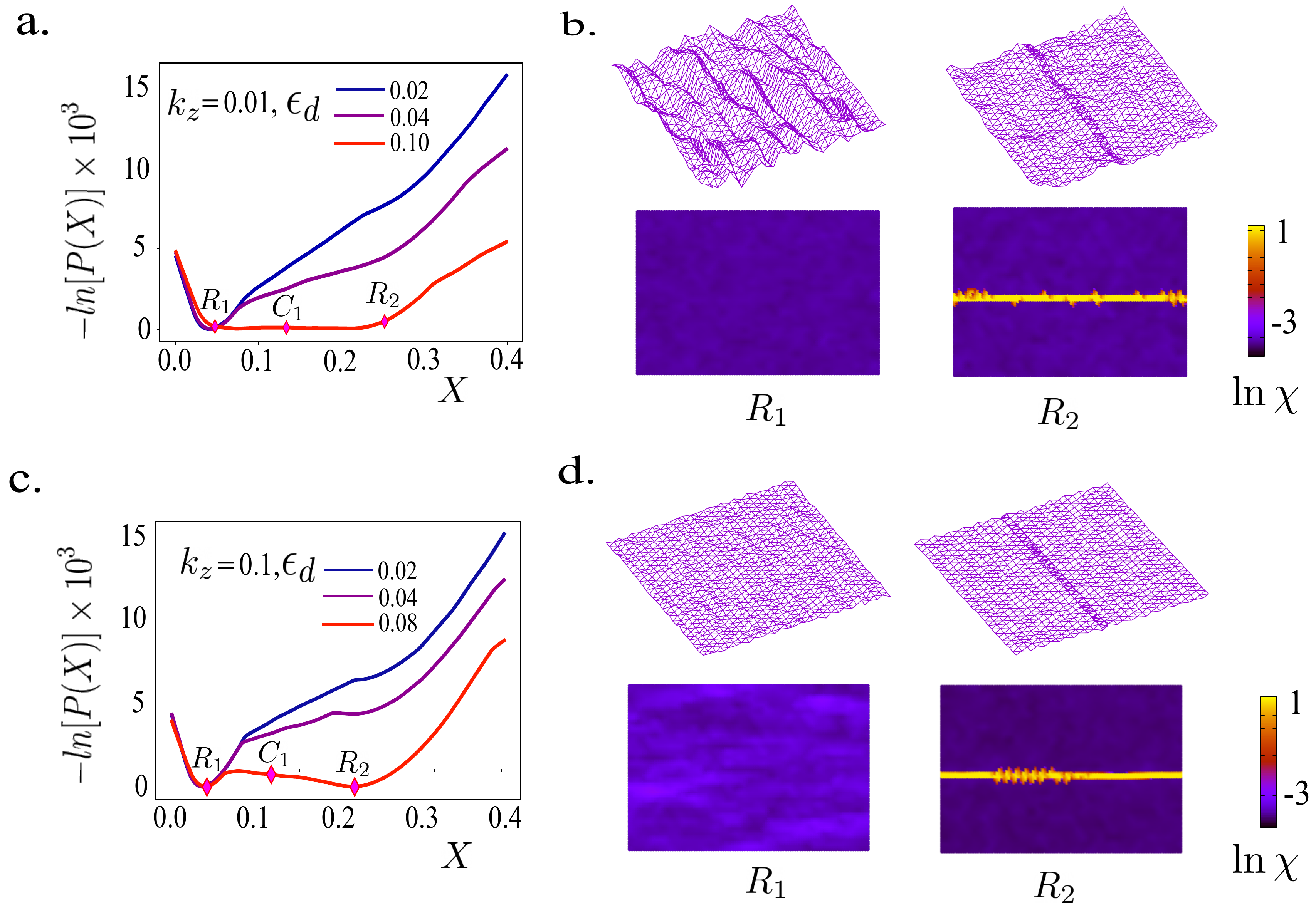}
\caption{ {\bf a.} Plot of effective free energy
$-\ln(P(X))$ obtained from SUS-MC of a $30\times30$ network with
$k_z = 0.01$ for ${\epsilon}_d = 0.02, 0.04, 0.10$. {\bf b.}
Configurations at $R_1$, $R_2$ for $k_z = 0.01$. {\bf c.} Same as
({\bf a.}) for $k_z = 0.1$ for ${\epsilon}_d = 0.02, 0.04, 0.08$.
{\bf d.} Configurations for $k_z = 0.1$ at $R_1$, $R_2$. In each
case, the first minimum is a ripple, and the second corresponds to
a configuration with a ripplocation. In {\bf b.} and {\bf d.}  the
upper panel is the averaged 3D configuration of the lattice and
the lower panel shows the 2D projections colored according to local
$\chi$ values in the $xy$ plane.} \label{ripplocation}
\end{figure*}
Figure \ref{wrinkle}{\bf a} shows the $\langle \Delta \rangle$
vs.~$\epsilon_d$ graph for different values of confinement. The
value of $\langle \Delta \rangle$ shows a finite jump for $k_z =
0.0, 0.01$ at $\epsilon_d = \epsilon_{w \to f}$
signifying the well-known ``wrinkle-to-fold transition''. For $k_z
= 0.0$ (see Fig.\ref{wrinkle} {\bf b}) and at a strain value
$\epsilon_d = 0.03$ below $\epsilon_{w \to f} = 0.08$, the network
is characterized by small $z$-fluctuations from wrinkles. For
$\epsilon_d  > \epsilon_{w \to f}$, large amplitude folds are
visible. Upon increasing confinement with $k_z =
0.01$, (Fig.\ref{wrinkle} {\bf c}) we observe that although single,
sharp, small wavelength peaks indicating folds are prevalent at
large strains $\epsilon_d > \epsilon_{w \to f} \approx 0.1$, the
amplitudes of these folds are smaller in comparison to those at
$k_z = 0.0$.  Note that even for very large strains, folds remain
single-valued in the $z$-direction, and the value of $X$ remains
small. For strong confinement, $k_z = 0.1$, a wrinkle-to-fold
transition is not observed (Fig \ref{wrinkle} {\bf d}): large folds
remain suppressed even at large $\epsilon_d$.

\subsection{\label{riptoriplo} Equilibrium ripple-to-ripplocation 
transition}
In this section, we describe the equilibrium thermodynamic phase
transition between the single-valued rippled phases and the
multi-valued ripplocated phase of the fluctuating confined sheet
at nonzero temperature.

In Fig.\ref{ripplocation} {\bf a}, we show $-\ln(P(X))$ for a
$30\times30$ network and $3$ different values of ${\epsilon}_d$
keeping $h_X = 0.0$. Here, we choose $k_z = 0.01$. The qualitative features in each of the $3$ cases are similar. The
first minimum $R_1$ at small $X$ corresponds to the ripple state.
The ripplocation phase $R_2$ has a large value of $X$. From
Fig.\ref{ripplocation} {\bf b} it is evident that the region around
the pleat has large local $\chi$ values. For larger values of $X$,
there exist higher order patterns corresponding to a sheet with
multiple ripplocations.  Here, we concentrate only on the transition
from a ripple state to one with a single ripplocation. On increasing
the external strain, the barrier height between the two minima of
the dimensionless free energy, $-\ln(P(X))$, decreases. For $\epsilon_d
= 0.10$, where the system is very close to the phase boundary, the
barrier height is about $\sim 10~k_{\rm B} T$. Figure
\ref{ripplocation}{\bf c} shows $-\ln(P(C))$ for a network with
$k_z = 0.1$. The free energy at three different values of ${\epsilon}_d
= 0.02$, 0.04, 0.08 and at $h_X = 0$ is qualitatively similar to
the $k_z = 0.01$ case. As before, the two minima at $R_1$ and $R_2$
represent the ripple and ripplocation phases, respectively. However,
coexistence is achieved at a smaller $\epsilon \sim 0.08$. The
barrier height near coexistence $\sim 80~k_{\rm B}T$ is significantly
higher compared to that at $k_z = 0.01$.

\begin{figure}[t!]
\includegraphics[scale=0.22]{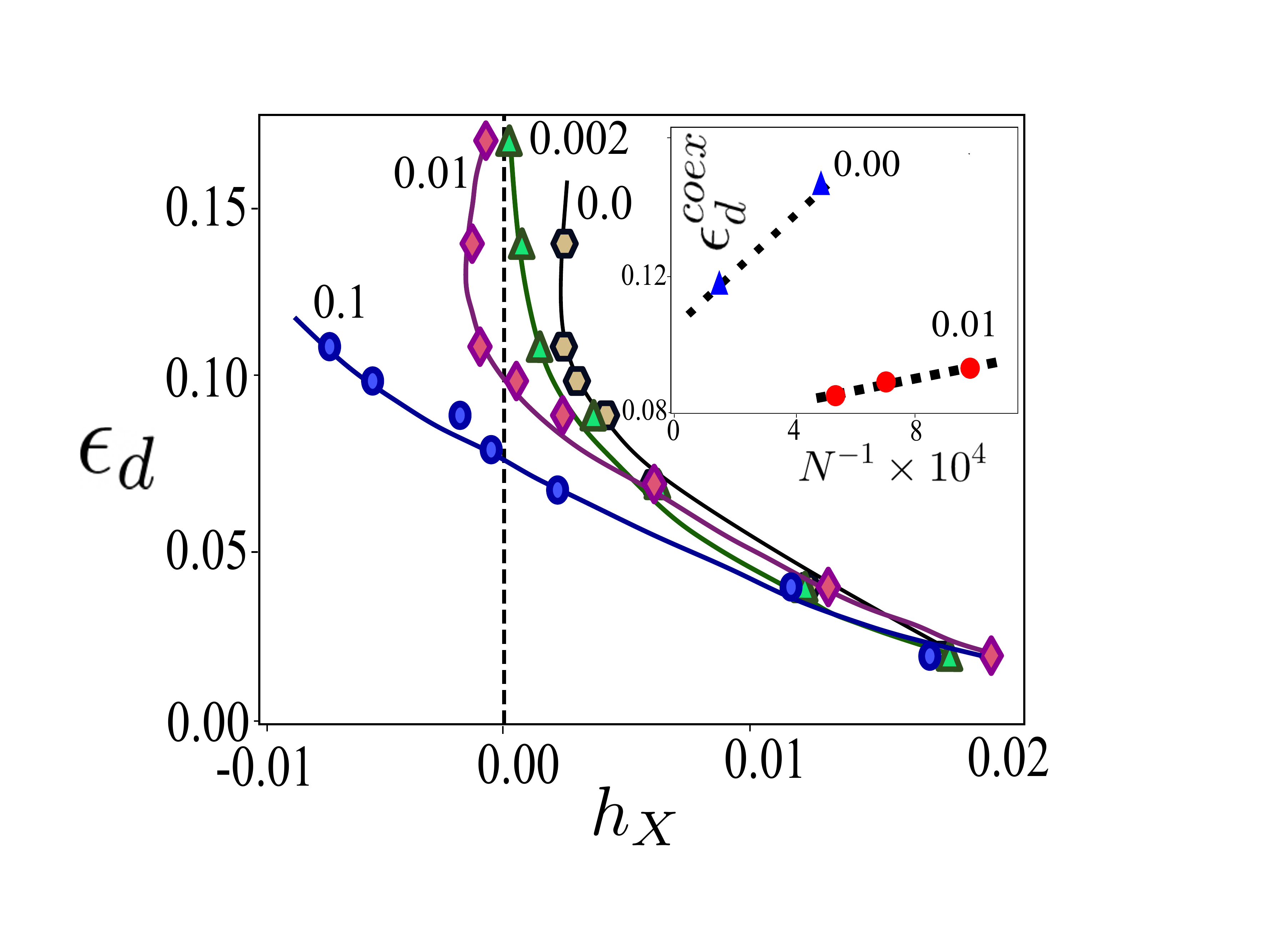}
\caption{Phase diagram of a $30\times30$ network in the
$h_X$-$\epsilon_d$ plane showing the evolution of the boundary
between the rippled and ripplocated phases as the strength of
confinement $k_z$ is varied. The numbers above the curves indicate the corresponding value for $k_z$. .Note that as $k_z$ is reduced, the
ripplocated phase ceases to exist at $h_X = 0$ for any strain. The inset
shows the coexistence point $\epsilon_d = \epsilon_{d}^{\rm coex}$
vs.~$N^{-1}$ at $h_X = 0$ for different values of $k_z$ as obtained
from SUS-MC simulations. The value of $\epsilon_{d}^{\rm coex}$
decreases gradually as we approach $N \to \infty$. \label{phase}}
\end{figure}
The results for the equilibrium transition can be summarized in the
phase diagram of Fig.\ref{phase}, where we have shown the equilibrium
phase boundaries for a $N = 30\times30$ system at non-zero temperature
for different confinement strengths $k_z$. With increasing confinement
$k_z$, the ripplocated phase at $h_X$ is achieved at a smaller value
of $\epsilon_d$. At the same time, as $k_z$ is reduced, we see that
the $\epsilon_d$ vs.~$h_X$ curve fails to intersect the $h_X = 0$
line, indicating that the ripplocated phase ceases to exist at any
strain for this case. Increasing $N$ shifts the
phase diagram downwards to smaller values of $\epsilon_d$, and one
obtains ripplocated phases at $h_X = 0$ where none existed at smaller
sizes. In the inset of Fig.\ref{phase}, we show that as $N \to
\infty$, the ripplocated phases appears even without confinement.
Note that the value of $\epsilon_d = \epsilon_d^{\rm coex}$ at $h_X
= 0$ decreases as one approaches $N \to \infty$.
Similarly, the free energy barrier between the
phases also has strong finite size effects.

It is clear from Fig.\ref{phase}{\bf a.} that
ripplocations either form at fixed $\epsilon_d$
by changing $h_X$ or at fixed $h_X$ by increasing $\epsilon_d$. The
latter protocol is analogous to the standard yielding transition
of solids at constant strain rate~\cite{pnas,
sas7}.

\begin{figure*}[t]
\includegraphics[scale=0.3]{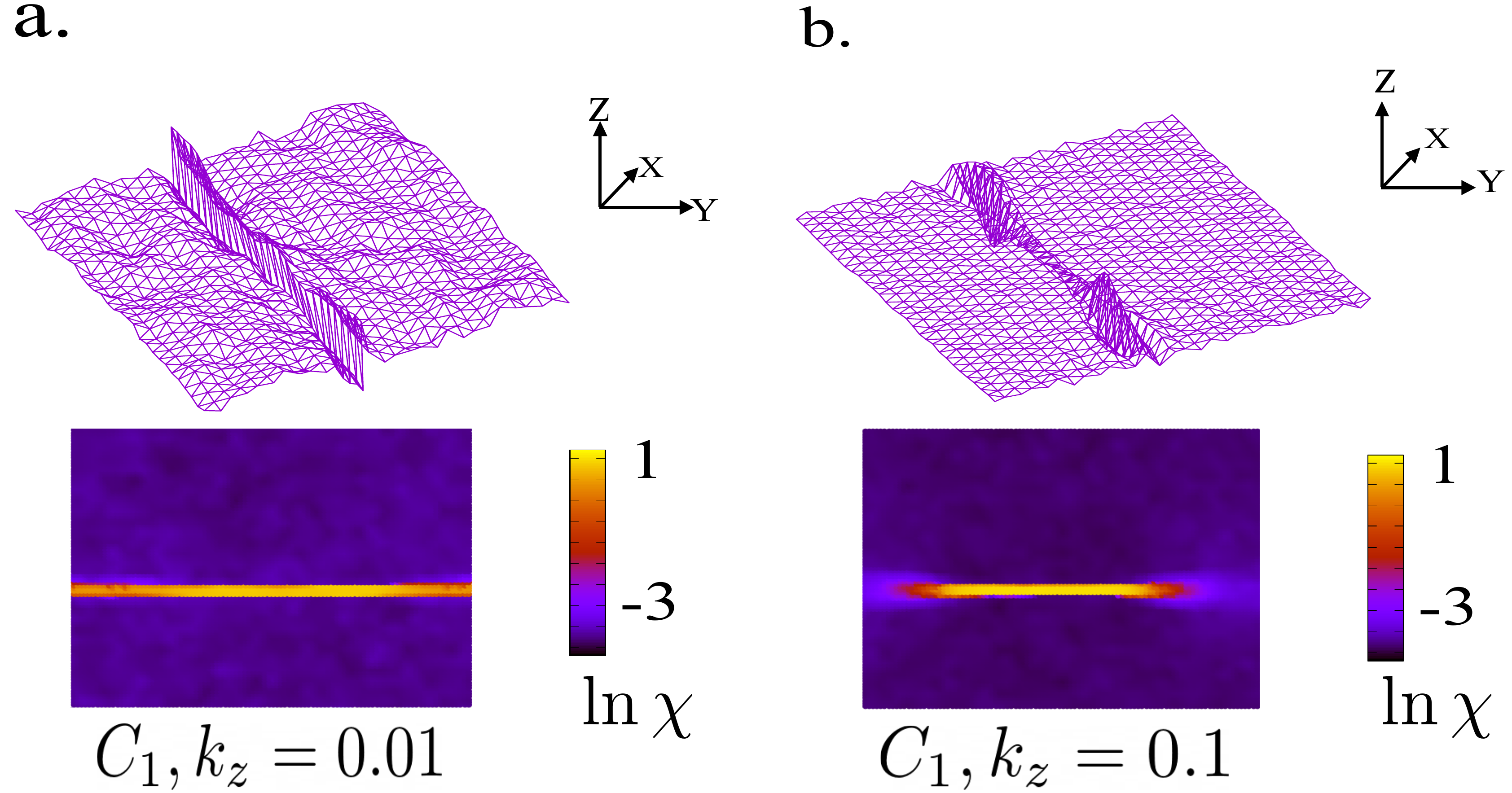}
\caption{Configurations at intermediate region
$C_1$ for a $30 \times 30$ network obtained from SUS-MC sampling
for {\bf a.} $k_z = 0.01$ (corresponding to
Fig.~\ref{ripplocation}a.) and {\bf b.} $k_z = 0.1$
(corresponding to Fig.\ref{ripplocation}c). The
upper panels show the averaged 3D configuration of the lattice and
the lower panel the local $\chi$ maps.} \label{intermediate}
\end{figure*}
\subsection{\label{sec:level2} Ripplocation precursors and 
intermediate structures}
An advantage of our SUS-MC sampling method is apparent from
Fig.\ref{intermediate}, where apart from the coexisting structures
at the two minima, we are also able to discover the intermediate
configurations along the transition path defined by the reaction
coordinate $X$. This is the least free energy equilibrium path
connecting the two phases. In Fig.\ref{intermediate} {\bf a} the
intermediate structure corresponding to $C_1$ (Fig.\ref{ripplocation}
{\bf a}) is shown. The ripplocation in this case is preceded by the
formation of a system-spanning fold. This is in sharp contrast with
the formation of pleats in 2D~\cite{sas7}, where the pleat forms
by a local transformation that produces a ``lip'' with two tips
where the displacement becomes singular.

The intermediate structure in Fig.\ref{intermediate} {\bf b} at
$C_1$ (see Fig.\ref{ripplocation} {\bf c}), on
the other hand, reveals that in this case indeed the ripplocation
is established by a local ``pinching'' where a lip with two tips
is formed (cf.~the point $O$ in Fig.\ref{schem1}{\bf b}).  This
pinching also results in the formation of a small bulge just near
the tip. The lip then extends all around the periodic boundary and
finally annihilates with itself once the ripplocation percolates
throughout the sheet. This similarity with pleat formation in a
flat 2D network can be explained by noting that with increasing
confinement strength $k_z$, formation of a bulk system-spanning
fold has a large energy cost; hence the network tends to remain
mostly in the 2D plane.

The presence or absence of a preceding wrinkle-to-fold transition
determines the intermediate configurations associated with the
ripple-to-ripplocation transition. This is illustrated in
Fig.\ref{wrinkle}{\bf a}.  For systems with strong confinement
above a certain threshold, where the wrinkle-to-fold transition is
suppressed, formation of bulk system-spanning folds is prohibited
(blue dot in Fig.\ref{wrinkle}{\bf a}); the pleat formation is
preceded by local pinching. Below that threshold, system spanning
folds are the precursors to ripplocation formation (green dot in
Fig.\ref{wrinkle}{\bf a}). The free energy barrier between a ripple
and a ripplocation is higher if an intermediate system spanning
fold does not exist.

\begin{figure}[t]
\includegraphics[scale=0.50]{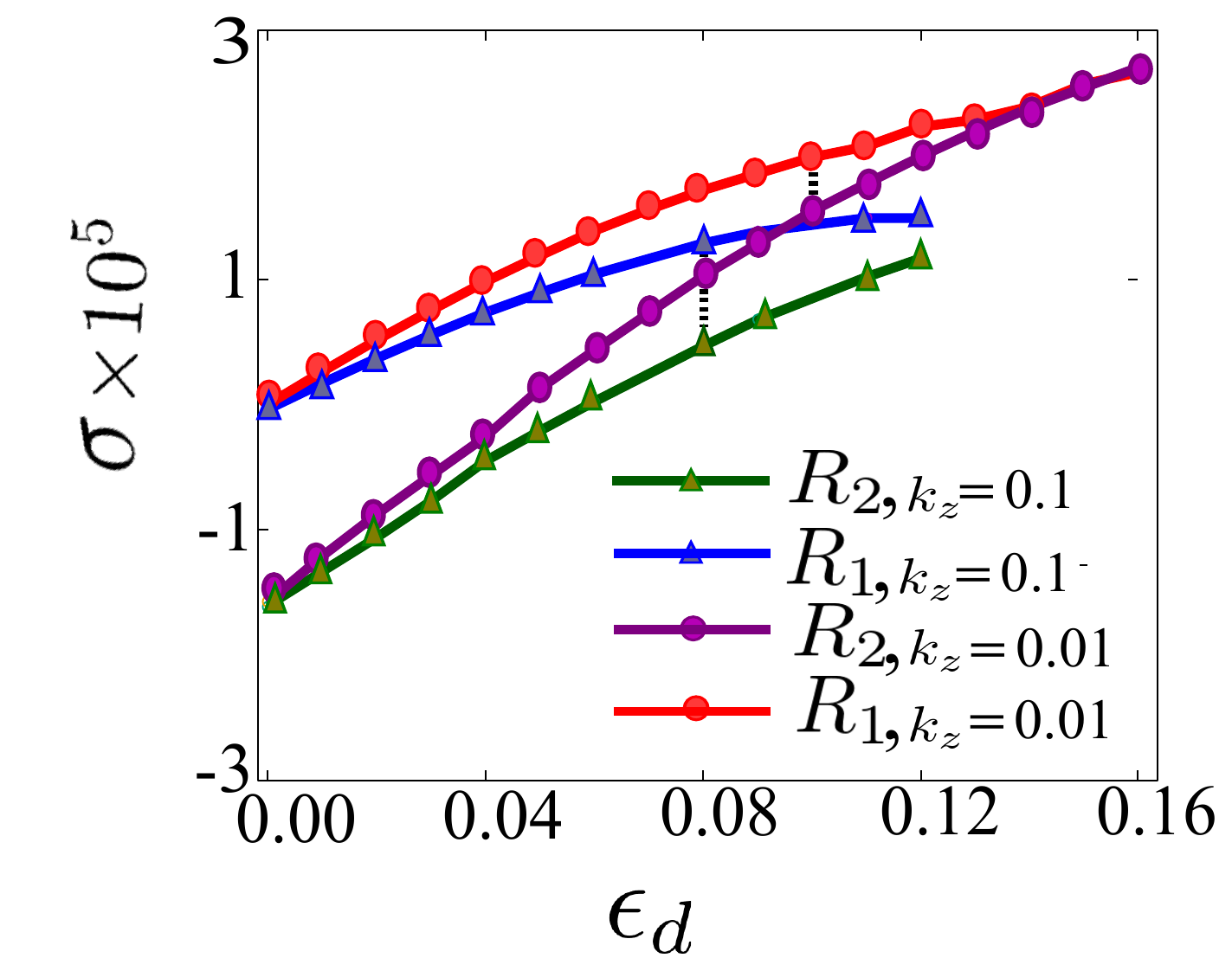}
\caption{Stress vs.~strain curves for rippled and ripplocated phases
of a $30\times30$ network. The nucleation of a ripplocated phase is always accompanied by a jump in stress (vertical dotted lines) as long as the strain is applied sufficiently slowly.} 
\label{strss}
\end{figure}
\subsection{\label{sec:level2} Mechanical signatures of the 
ripple-to-ripplocation transition}
As mentioned earlier, ripples and ripplocations are separated by
high barriers. Signatures of such high barrier values are apparent
when we compute the stress for the ripple ($R_1$) and ripplocation
($R_2$) phases. This is illustrated in Fig.\ref{strss}, where we
have plotted curves of conjugate stress $\sigma$ vs.~$\epsilon_d$
for ripples and ripplocated phases at two different values of $k_z$
as obtained from SUS-MC calculations.  We obtain $\sigma$ in the standard way~\cite{ums} by averaging the corresponding virial over a slab of thickness $Z_w$ centered about the plane $z=0$  wholly containing the reference lattice points.  
Since the value of $\sigma$ is computed for a possibly metastable phase, the rate of deformation
$\dot{\epsilon}$ plays an important role.

For large $\dot{\epsilon}$ the transition happens at the limit of
metastability of the rippled (wrinkled or folded) phase. For
quasistatic deformation, on the other hand, the transition occurs
at the equilibrium phase boundary.  At any intermediate deformation
rate, the transition should occur somewhere in between these limits.
The ripplocation phase, which localizes stress within the ripplocation
while relieving it elsewhere~\cite{sas7} has a relatively lower
value of average stress. The formation of the ripplocated phase is
therefore always expected to be associated with a jump in the value
of stress.

The formation of ripplocation is an irreversible deformation; once
the network attains a ripplocation state it is unable to revert out
of this due to high barriers. The jump in stress at the transition
follows the barrier height between two phases; as we increase the
confinement $k_z$, the magnitude of the jump in stress increases.
This offers a unique way of distinguishing the formation of
ripplocations from either system spanning folds or localized wrinkles
in realistic systems by subjecting them to deformation experiments.
Depending on whether a large or small jump is seen when structures
are deformed at equal rates, one should be able to extract the
barriers and estimate the magnitude of the confining interlayer
potential.  Small barriers point to a transition from a fold to a
ripplocation while large barriers are associated with the formation
of ripplocations directly from wrinkles. The
deformation behavior of these materials is therefore highly
non-linear~\cite{barsoum2003, 1barsoum2003, barsoum2004, 1barsoum2004,
barsoum2005, barsoum2013, kushima2015, gruber2016, Tucker, Freiburg}.

Unravelling the dynamics of this transition and predicting the yield
point, i.e.\ the deformation at which the stress decreases, which is
necessarily a function of the strain rate, requires more intense
study; work in this direction is planned for the near future~\cite{pnas,
reddy}.

\section{\label{sec:level1} Summary and outlook}
In this paper, we have described in detail the $T > 0$ pleating
transition and formation of a ripplocation in a confined 2D sheet
with out-of-plane fluctuations when deformed by a pure shear. Our
motivation is to understand how plastic deformation occurs in a
system where defects or atomic rearrangements are not possible.
Similar to our work on a ``ghost network''~\cite{sas4,sas7}, we
show the existence of the ripple-to-ripplocation transition when
such a sheet is confined by parallel walls but fluctuations in the
perpendicular direction are allowed. The ripple phase evolves from
wrinkles characterized by small height fluctuations at small strain
to large amplitude folds at high strain values. While such
wrinkle-to-fold transitions have been reported for a variety of
situations in 2D membrane networks, ripplocations are a novel type
of structure not identified in any of these earlier works~\cite{Milner,
Cerda-Maha, Zhang, Brau, Li-wrinkle} where the surface is taken to
be always single-valued. Establishing the existence of such
multi-valued deformation structures as a consequence of a first-order
pleating phase transition is a primary contribution of our work.
In order to describe the transition we define an external field
conjugate to the non-affine displacements $X$.The thermodynamic
variable $X$ behaves as a reaction coordinate to describe our
ripple-to-ripplocation transition.

The main conclusions of this work can be summarized as follows:
\begin{enumerate}
\item The response to deformation in our model is characterized by
a first-order phase transition. The ripple phase and the ripplocation
phase are separated by large free energy barriers. The free energy
barrier between ripple and ripplocation strongly depends on the
strength of the confinement.
\item Apart from this pleating transition, the wrinkle-to-fold
transition is also present in our model. This transition influences
the intermediate structures in the ripple-to-ripplocation transition.
\item The ripple to ripplocation transition occurs at smaller strains for higher values of confinement.
\item Two different types of ripplocation formation can be seen in
our calculations, depending on the strength of
the confinement. For weak confinement, the ripplocation is preceded
by bulk system-spanning ripples. In case of stronger confinement,
a pleat tip is formed, which percolates through the system to form
ripplocations.
\end{enumerate}

Finally, a word of caution. It should be pointed
out that throughout this work, we distinguish between the terms
ripple and ripplocation, depending on whether the height variable
is single- or multi-valued, respectively. This may be contrasted
with the term "ripplocation" used in recent literature \cite{kushima2015,
gruber2016}, which referred to any large localized variation of
height.

Our model for a confined crystalline sheet could
be modified in many ways, such that it becomes more realistic. Real confined membranes, e.g., do have an intrinsic curvatures and bending
rigidity. This is also true for real layered
solids.  Such effects have been neglected in our conceptual model.
However, regardless of the minute details, we believe that essential
qualitative aspects of the pleating transition like the stress-strain
curve, the finite-size effects and the intermediate structures will
be similar to those described in this paper. We are in the process
of extending our work to a realistic model of graphene where some
of these questions can be addressed and the effect of ripplocation deformation on physical properties can be
explored ~\cite{sandhya}.

The exact nature of the phases is also strongly dependent on the
dynamics of external loading. Dynamical effects have been neglected
in the present work and their incorporation will be an interesting
exercise once accurate experimental results are available for
comparison.

\acknowledgements
 This project was funded by intramural funds at TIFR Hyderabad from
the Department of Atomic Energy (DAE), India. TS-D would like to
thank Department of Science and Technology, India, for funding.


\begin{thebibliography}{}
%
\bibitem{CL} 
P. Chaikin and T. Lubensky, 
{\it Principles of Condensed Matter Physics} 
(Cambridge University Press, Cambridge, 1995).
%
\bibitem{Landau} 
L. Landau and E. M. Lifshitz, 
{\it Theory of Elasticity, 3rd ed.}
(Pergamon, New York, 1986).
%
\bibitem{rob}
R. Phillips,
{\it Crystals, defects and microstructures: Modeling across scales} 
(Cambridge University Press, Cambridge, 2004).
%
\bibitem{hirth}
J. P. Hirth and J. Lothe, 
{\it Theory of Dislocations} 
(McGraw-Hill, New York, 1967).
%
\bibitem{nabarro} 
F. R. N. Nabarro and M. S. Duesbery, Eds., 
{\it Dislocations in Solids, Vol. 10}  
(Elsevier, Amsterdam, 1996), p.505-594.
%
\bibitem{hierarchy} 
H. Vandeparre, M. Pineirua, F. Brau, B. Roman, J. Bico, C. Gay, W. Bao, 
C. N. Lau, P. M. Reis, and P. Damman, 
Phys. Rev. Lett. {\bf 106}, 224301 (2011).
%
\bibitem{Milner} 
S. T. Milner, J.-F. Joanny, and P. Pincus, 
Europhys. Lett. {\bf 9}, 495 (1989).
%
\bibitem{Cerda-Maha} 
E. Cerda and L. Mahadevan, 
Phys. Rev. Lett. {\bf 90}, 074302 (2003).
%
\bibitem{Zhang} 
Q. Zhang and T. A. Witten, 
Phys. Rev. E {\bf 76}, 041608 (2007).
%
\bibitem{Brau} 
F. Brau, H. Vandeparre, A. Sabbah, C. Poulard, A. Boudaoud, and P. Damman, 
Nat. Phys. {\bf 7}, 56 (2011).
%
\bibitem{Li-wrinkle}
B. Li, Y.-P. Cao, X.-Q. Feng, and H. Gao, 
Soft Matter {\bf 8}, 5728 (2012).
%
\bibitem{sas4}
S. Ganguly, P. Nath, J. Horbach, P. Sollich, S. Karmakar, and S. Sengupta, 
J. Chem. Phys. {\bf 146}, 124501 (2017).
%
\bibitem{sas7} 
S. Ganguly, D. Das, J. Horbach, P. Sollich, S. Karmakar, and S. Sengupta,
J. Chem. Phys. {\bf 149}, 184503 (2018).
%
\bibitem{spectrin1}
D. H. Boal, U. Seifert, and A. Zilker,  
Phys. Rev. Lett. {\bf 69}, 3405 (1992).
%
\bibitem{spectrin2}
H. Li and G. Lykotrafitis, 
Biophys. J. {\bf 102}, 75 (2012).
%
\bibitem{muthu}
M. Muthukumar, C. K. Ober, and E. L. Thomas, 
Science {\bf 277}, 1225 (1997).
%
\bibitem{origami}
L. Mahadevan and S. Rica, 
Science {\bf 307}, 1740 (2005).
%
\bibitem{kirigami1}
D. M. Sussman, Y. Cho, T. Castle, X. Gong, E. Jung, S. Yang, and R. D. Kamien, 
Proc. Natl. Acad. Sci. USA {\bf 112}, 7449 (2015).
%
\bibitem{kirigami2}
T. Castle,Y. Cho, X. Gong, E. Jung, D. M. Sussman, S. Yang, and R. D. Kamien, 
Phys. Rev. Lett. {\bf 113}, 245502 (2014).
%
\bibitem{irvine}
W. T. M. Irvine, V. Vitelli, and P. M. Chaikin, 
Nature {\bf 468}, 947 (2010).
%
\bibitem{grason1}
G. M. Grason and B. Davidovitch, 
Proc. Natl. Acad. Sci. USA {\bf 110}, 12893 (2013).
%
\bibitem{grason2}
A. Azadi and G. M. Grason, 
Phys. Rev. Lett. {\bf 112}, 225502 (2014).
%
\bibitem{grason3}
A. Azadi and G. M. Grason, 
Phys. Rev. E {\bf 94}, 013003 (2016).
%
\bibitem{narayanan1}
H. King, R. D. Schroll, B. Davidovitch, and N. Menon, 
Proc. Natl. Acad. Sci. USA {\bf 109}, 9716 (2012).
%
\bibitem{narayanan2}
A. D. Cambou and N. Menon, 
Proc. Natl. Acad. Sci. USA {\bf 108}, 14741 (2011).
%
\bibitem{flat-fold1}
L. H. Dudte, E. Vouga, T. Tachi, and L. Mahadevan, 
Nat. Mater. {\bf 15}, 583 (2016).
%
\bibitem{flat-fold2}
J. L. Silverberg, J-H. Na, A. A. Evans, B. Liu, T. C. Hull, C. D. Santangelo, 
R. J. Lang, R. C. Hayward, and I. Cohen, 
Nat. Mater. {\bf 14}, 389 (2015).
%
\bibitem{frank} 
F. C. Frank and A. N. Stroh, 
Proc. Phys. Soc. London {\bf 65}, 811 (1952).
%
\bibitem{barsoum2003}
M. W. Barsoum, T. Zhen, S. R. Kalidindi, M. Radovic, and A. Murugaiah, 
Nat. Mater. {\bf 2}, 107 (2003).
%
\bibitem{1barsoum2003}
A. Murugaiah, M. W. Barsoum, S. R. Kalidindi, and T. Zhen, 
J. Mater. Res. {\bf 19}, 1139 (2004).
%
\bibitem{barsoum2004}
M. W. Barsoum, A. Murugaiah, S. R. Kalidindi, and T. Zhen, 
Phys. Rev. Lett. {\bf 92}, 255508 (2004).
%
\bibitem{1barsoum2004}
M. W. Barsoum, A. Murugaiah, S. R. Kalidindi, T. Zhen, and Y. Gogotsi, 
Carbon {\bf 42}, 1435 (2004).
%
\bibitem{barsoum2005} 
M. W. Barsoum, T. Zhen, A. Zhou, S. Basu, and S. R. Kalidindi,
Phys. Rev. B {\bf 71}, 134101 (2005).
%
\bibitem{barsoum2013}
M. W. Barsoum,
{\it MAX Phases: Properties of Machinable Ternary Carbides and Nitrides}
(Wiley-VCH, Weinheim, 2013).
%
\bibitem{kushima2015}
A. Kushima, X. Qian, P. Zhao, S. Zhang, and J. Li,
Nano Lett. {\bf 15}, 1302 (2015).
%
\bibitem{gruber2016}
J. Gruber, A. C. Lang, J. Griggs, M. L. Taheri, G. J. Tucker, and M. W. Barsoum,
Sci. Rep. {\bf 6}, 33451 (2016).
%
\bibitem{Tucker}
M. W. Barsoum and G. J. Tucker,
Scr. Mater. {\bf 139}, 166 (2017).
%
\bibitem{Freiburg}
D. Freiberg, M. W. Barsoum, and G. J. Tucker,
Phys. Rev. Mater. {\bf 2}, 053602 (2019).
%
\bibitem{Barsoum-Zhao}
M. W. Barsoum, X. Zhao, S. Shanazarov, A. Romanchuk, S. Koumlis, 
S. J. Pagano, L. Lamberson, and G. J. Tucker,
Phys. Rev. Mater. {\bf 3}, 013602 (2019). 
%
\bibitem{Aslin2019}
J. Aslin, E. Mariani, K. Dawson, and M. W. Barsoum,
Nat. Commun. {\bf 10}, 686 (2019).
%
\bibitem{falk}
M. L. Falk and J. S. Langer,  
Phys. Rev. E {\bf 57}, 7192 (1998).
%
\bibitem{sas1}
S. Ganguly, S. Sengupta, P. Sollich, and M. Rao, 
Phys. Rev. E {\bf 87}, 042801 (2013).
%
\bibitem{popli} 
P. Popli, S. Kayal, P. Sollich, and S. Sengupta, 
Phys. Rev. E {\bf 100}, 033002 (2019).
%
\bibitem{pnas}
P. Nath, S. Ganguly, J. Horbach, P. Sollich, S. Karmakar, and S. Sengupta, 
Proc. Natl. Acad. Sci. USA {\bf 115}, E4322 (2018).
%
\bibitem{protein} 
D. D. Prakashchand, N. Ahalawat, H. Khandelia, J. Mondal, and S. Sengupta, 
PLoS Comput. Biol. {\bf 15}, e1006665 (2019).
%
\bibitem{sas2}
S. Ganguly, S. Sengupta, and P. Sollich, 
Soft Matter {\bf 11}, 4517 (2015).
%
\bibitem{poplitrap} 
P. Popli, S. Ganguly, and S. Sengupta, 
Soft Matter {\bf 14}, 104 (2018).
%
\bibitem{SUS}
P. Virnau and M. M\"uller,
J. Chem. Phys. {\bf 120}, 10925 (2004).
%
\bibitem{binder}
K. Binder and D. W. Heermann,
{\it Monte Carlo Simulation in Statistical Physics: An Introduction, Fifth Edition}
(Springer, New York, 2010).
%
\bibitem{ums}
D. Frenkel and B. Smit,
{\it Understanding Molecular Simulation: From Algorithms to Applications}
(Academic Press, San Diego, 2002).
%
\bibitem{borgs90}
C. Borgs and R. Koteck\'y,
J. Stat. Phys. {\bf 61}, 79 (1998).
%
\bibitem{ferrenberg88}
A. M. Ferrenberg and R. H. Swendsen,
Phys. Rev. Lett. {\bf 61}, 2635 (1988).
%
\bibitem{reddy} 
V. S. Reddy, P. Nath, J. Horbach, P. Sollich, and S. Sengupta, 
Phys. Rev. Lett. {\bf 124}, 025503 (2020).
%
\bibitem{schuster1980}
J. C. Schuster, H. Nowotny, and C. Vaccaro,
J. Solid State Chem. {\bf 32}, 213 (1980).
%
\bibitem{sandhya} 
D. Das, C. Sandhya, J. Horbach, P. Sollich, T. Saha-Dasgupta, and S. Sengupta 
(manuscript under preparation). 
%
%
%
%
%
%
%
%
%
%
%

\end{thebibliography}
\end{document}